\documentclass[8pt]{article}  \usepackage{times}
\usepackage{graphicx}

\usepackage{color}

\usepackage[round,numbers,sort&compress]{natbib} 
\begin{document}

\twocolumn[{\LARGE \textbf{The capacitance and electromechanical coupling of lipid membranes close to transitions. The effect of electrostriction.\\*[0.2cm]}}
{\large Thomas Heimburg$^{\ast}$\\*[0.1cm]
{\small $^1$Niels Bohr Institute, University of Copenhagen, Blegdamsvej 17, 2100 Copenhagen \O, Denmark}\\

{\normalsize ABSTRACT\hspace{0.5cm} Biomembranes are thin capacitors with the unique feature of displaying phase transitions in a physiologically relevant regime. We investigate the voltage and lateral pressure dependence of their capacitance close to their chain melting transition.  Since the gel and the fluid membrane have different area and thickness, the capacitance of the two membrane phases is different.  In the presence of external fields, charges exert forces that can influence the state of the membrane, thereby influencing the transition temperature. This phenomenon is called electrostriction. We show that this effect allows us to introduce a capacitive susceptibility that assumes a maximum in the melting transition with an associated excess charge. As a consequence, there exist voltage regimes where a small change in voltage can lead to a large uptake of charge and a large capacitive current.  Furthermore, we consider electromechanical behavior such as pressure-induced changes in capacitance, and the application of such concepts in biology.\\*[0.0cm] }}
]

\noindent\footnotesize {$^{\ast}$corresponding author, theimbu@nbi.dk}\\

\footnotesize{\textbf{Abbreviations:} DMPC - 1,2 dimyristoyl-3-sn-phosphatidylcholine; DMPG - 1,2 dimyristoyl-3-sn-phosphatidylglycerol; AITC - allylisothiocyanate ; } 
\footnotesize{pdf - probability distribution function; HEK cell - human embryonic kidney cell; TRP channel - transient receptor potential channel; ADPR - adenosine diphosphate ribose}

\normalsize
\section*{Introduction}
Biological membranes provide a barrier between cells and organelles that serves to maintain differences in chemical and electrical potentials by separating molecules and ions.  The electrical phenomena which result from transient changes in these electrochemical potentials provide the basis for our present understanding of the electrophysiology of biomembranes \cite{Hille1992, Johnston1995}.  The cell membrane consists mainly of a lipid bilayer into which proteins are embedded.  It is widely believed that the lipid bilayer itself is impermeable to water, ions and molecules. Therefore, electrophysiology considers the membrane to be a capacitor, and ion channel proteins are regarded as electrical resistors. The nerve pulse, for instance, is considered as a propagating segment of charged capacitor loaded by currents through the channel proteins \cite{Hodgkin1952b}. 


The capacitance, $C_m$, defines how much charge, $q$, is stored on two capacitor plates at a fixed voltage, $V_m$,
\begin{equation}
\label{eq:Intro.01}
q=C_m\cdot V_m \; .
\end{equation}
For a parallel plate capacitor, $C_m$ is given by
\begin{equation}
\label{eq:Intro.02}
C_m=\epsilon_0\epsilon\frac{A}{D} \;,
\end{equation}
with vacuum permittivity $\epsilon_0=8.854á10^{-12}$\,F/m and dielectric constant $\epsilon\approx 2-4$. $A$ is the area of the membrane, and $D$ is its thickness. 

Excitatory processes in cells are typically accompanied by changes in voltage. During nerve pulses, for instance, the voltage changes transiently by about 100 mV in one millisecond.  Voltages as large as 100 mV are also typical in the voltage clamp experiments that are used to measure protein conductances \cite{Hodgkin1952a}. In electroporation experiments, used to transport drugs or DNA into cells, voltages can be on the order of several 100 mV  \cite{Neumann1999, Gehl2003}. Stimulation voltages for nerve pulses can be of order up to 1\,V (e.g., \cite{Kassahun2010}). 

A change in voltage leads to a capacitive current because the charge on the capacitor changes. The capacitive current induced by a change in voltage is given as
\begin{equation}
\label{eq:Intro.03}
\frac{dq}{dt}=\frac{d}{dt}\left(C_m\cdot V_m\right)=C_m\frac{dV_m}{dt}+ V_m\frac{dC_m}{dt}
\end{equation}
In electrophysiological models such as the Hodgkin-Huxley model for nerve pulse propagation and in the interpretation of voltage clamp experiments, it is assumed that the capacitance of biomembranes (in particular of nerves) is independent of voltage (i.e., $C_m$ is constant) so 
that the second term on the right of this equation is zero (e.g., \cite{Hodgkin1952b, Johnston1995}).  This is equivalent to assuming that 
membrane dimensions are unaffected by electrical phenomena and that the excitation of membranes does not change their dimensions.  
The second of these assumptions is known to be incorrect because changes in the thickness of nerve membranes during the action potential have been observed \cite{Iwasa1980a, Iwasa1980b, Tasaki1980, Tasaki1982b, Tasaki1989, Heimburg2005c, Kim2007}. Furthermore, there are numerous reports in the literature on voltage induced changes in membrane bending, i.e., caused by flexoelectricity or mechanoelectricity \cite{Ochs1974, Raphael2000, Petrov2006}. In the recent years, we have proposed that the voltage changes during the nerve pulse are actually related to changes in capacitance \cite{Heimburg2005c, Heimburg2007b, Heimburg2008, Andersen2009, Villagran2011, Lautrup2011}.\\

In this theoretical paper we show that the assumption of constant capacitance is incorrect, especially if one is close to chain melting transitions in the lipid membrane. Biological membranes display transitions close to physiological temperature. Heat capacity maxima are typically found 10-15 degrees below physiological or growth temperature, both for bacterial membranes from \textit{E.coli} and bacillus subtilis, for lung surfactant \cite{Heimburg2005c} and for nerves from the spine of rats (S. B. Madsen, N. V. Olsen, unpublished data). The fluid state membrane is thinner than the gel membrane. Simultaneously, the area of the fluid membrane is larger \cite{Heimburg1998}. This implies that the capacitance of the fluid membrane is larger than that of the gel membrane. Thus, any phenomenon in the biomembrane related to a phase transition will influence its capacitance. For instance, the temperature of the phase transition is influenced by charges on the capacitor because electrostatic forces act on the capacitor plates. Further, hydrostatic pressure \cite{Ebel2001}, lateral pressure \cite{Heimburg2005c} or the addition of drugs like anesthetics \cite{Heimburg2007c} influence the position of the melting transition. Therefore, the capacitance must be also a function of voltage, hydrostatic pressure, lateral pressure, and the concentration of anesthetics. 

We show here that close to a transition voltage is able to change the dimensions of a membrane and the capacitance.  We also show how the electrical properties of the membrane are affected by the application of lateral pressure or tension in a membrane. This gives rise to pressure-induced capacitive currents or pressure-induced voltage changes.  While we restrict ourselves primarily to the phenomenon of electrostriction, i.e., the force that capacitive charges exert on the capacitor, we also discuss polarization effects.



\section*{Theory}\label{Theory}

Consider a capacitor whose equilibrium properties depend on voltage $V_m$ only, i.e., all other intensive variables of the system such as pressure and temperature are kept constant. We write
\begin{equation}
\label{eq:T.01}
dq=\left(\frac{\partial q}{\partial V_m}\right)dV_m\equiv \hat{C}_m dV_m \;.
\end{equation}
Here we introduce the function $\hat{C}_m=(\partial q/\partial V_m)$, which we call the capacitive susceptibility. Note, that the definition of $\hat{C}_m$ differs from that of the capacitance that is given by $C_m=q/V_m$.
According to eq. (\ref{eq:Intro.01}), eq. (\ref{eq:T.01}) corresponds to
\begin{eqnarray}
\label{eq:T.02}
dq&=&\left(\frac{\partial (C_m V_m)}{\partial V_m}\right)dV_m=\left(C_m+V_m\frac{\partial C_m}{\partial V_m}\right)dV_m \nonumber\\
&=&C_m dV_m+V_m dC_m \;.
\end{eqnarray}
This equation takes into account that the changes of the charge on a capacitor are not solely due to voltage changes but also to voltage-induced changes in capacitance, which are described by the capacitive susceptibility
\begin{equation}
\label{eq:T.03}
\hat{C}_m \equiv C_m+V_m\frac{\partial C_m}{\partial V_m} \;.
\end{equation}
If the capacitance $C_m$ is independent of voltage, we obtain $\hat{C}_m=C_m$. However, the last term on the right hand side of eq. (\ref{eq:T.03}) can become large close to transitions in biomembranes as we will show below. In the context of transitions, this term can be considered an excess capacitance. It is proportional to the voltage. Therefore it is zero at zero voltage where the capacitor is not charged. In contrast, the function $C_m$ has a finite value since it depends only on the dimensions of the capacitor. 

The capacitive susceptibility  $\hat{C}_m$ is fully analogous to other susceptibilities such as heat capacity, $(dH/dT)_p$, isothermal volume compressibility, $-(dV/dp)_T$, and isothermal area compressibility, $-(dA/d\Pi)_T$. The capacitive susceptibility has been used before, e.g., by \cite{Carius1976}.

\subsection*{Electrostriction}\label{Theory_1}
Electrostriction is the generation of a mechanical force on a capacitor by the electrostatic attraction of the charges on the two capacitor plates. In the absence of other forces, increasing the voltage on a capacitor will generally tend to deform the capacitor such that its thickness is reduced. \\

Consider a membrane with fixed thickness $D$, a capacitance $C_m$ (given by eq. (\ref{eq:Intro.02})) and a transmembrane voltage, $V_m$.
The field across the membrane is $E=V_m/D$ (assuming a uniform dielectric constant in the membrane interior), and the charge is $q=C_m\;V_m$. The force, $\mathcal{F}$, acting on this capacitor is given by \cite{Vanselow1966a}
\begin{equation}
\label{eq:T_1.01}
\mathcal{F}=\frac{1}{2}E\cdot q= \frac{1}{2}\frac{V_m}{D}q = \frac{1}{2}\frac{C_m\;V_m^2}{D} \;.
\end{equation}
The force $\mathcal F$ and the field $E$ are vectors normal to the membrane surface.  Assuming that the fluid membrane has a capacitance of $C_m=0.5$ $\mu$F/cm$^2$ = $0.5\cdot 10^{-2}$ F/m$^2$, this results in a pressure on the membrane of $p=10^4$\,N/m$^2$ at 100 mV, which corresponds to 0.1 bar. Due to its quadratic dependence on voltage, this pressure is 100 times larger ($p=10$ bar) at $V_m=1$ volt. In contrast to hydrostatic pressure this pressure has a direction normal to the membrane. This implies that increasing this force results in a reduction of thickness and an increase in area. Since the melting of a membrane is linked to an increase in area, an increase of transmembrane voltage can therefore potentially melt a membrane.

At constant voltage, the work done by the electrical field upon melting of the membrane  is given by 
\begin{equation}
\label{eq:T_1.02}
\Delta W_c=\int_{D_g}^{D_f}\mathcal{F}dD=\frac{1}{2} \epsilon_0 V_m^2\int_{D_g}^{D_f} \epsilon\frac{A}{D^2}dD \;,
\end{equation}
where $D_g$ and $D_f$ are the thickness of the gel and the fluid membrane, respectively. For constant area this relation yields the familiar expression for the work done on a capacitor, $\frac{1}{2} C_m V_m^2$. However, the membrane area does not stay constant here.

The area in the fluid state of DPPC is 24.6\% larger than in the gel state and the thickness is 16.3\% smaller \cite{Heimburg1998}. If we assume a dielectric constant $\epsilon$ independent of voltage and membrane state, this difference is 
\begin{equation}
\label{eq:T_1.05}
C_m^{fluid}=\epsilon_0\epsilon \frac{A_{g}\cdot(1+0.246)}{D_{g}\cdot(1-0.163)}=1.49\; C_m^{gel} \;.
\end{equation}
Thus, the capacitance in the fluid phase is about 1.5 times larger than that of the gel phase, and the force across the membrane is 1.78 time larger than in the gel phase. In the presence of a voltage difference, a sudden change in membrane state can lead to significant capacitive currents (see discussion).

The assumption made above of a voltage-  and temperature-independent dielectric constant may not be correct. Paraffin oil has a dielectric constant of $2.2-4.7$ and olive oil has $3.1$ while paraffin wax has $2.1-2.5$ \cite{ASI}. Therefore, it may be possible that the dielectric constant in the fluid membrane is somewhat higher.  Lacking reliable data for membranes, we consider $\epsilon$ to be constant.

\subsection*{Voltage dependence of the melting temperature}\label{Theory_2}
In the following we describe the melting of a membrane in the presence of a transmembrane voltage. We approximate the excess heat capacity in the absence of voltage, $\Delta c_{p,0} =d(\Delta H_0(T))/dT$, by assuming that the transition is governed by a van't Hoff law, i.e., by a two state-transition from gel to fluid with a temperature dependent equilibrium constant $K(T)$. The temperature dependence of the excess enthalpy, $\Delta H_0(T)$ is given by
\begin{eqnarray}
	\label{eq:T_2.01}
	\Delta H_0(T)=\frac{K(T)}{1+K(T)}\Delta H_0 \qquad \mbox{with}\qquad \nonumber\\
	K(T)=\exp{\left(-n\cdot\frac{\Delta H_0}{k}\left(\frac{1}{T}-\frac{1}{T_m}\right)\right)} \;,
\end{eqnarray}
where $n$ is a cooperative unit size that determines how many lipids undergo a transition at the same time. For DPPC, the total excess enthalpy of the transition is $\Delta H_0=39$ kJ/mol, and the melting temperature is $T_m=314.2$ K. The transition of DPPC unilamellar vesicles is reasonably well described by $n=100$. (For details of this calculation see \cite{Heimburg2007a}.)
\begin{figure}[ht!]
    \begin{center}
	\includegraphics[width=8.5cm]{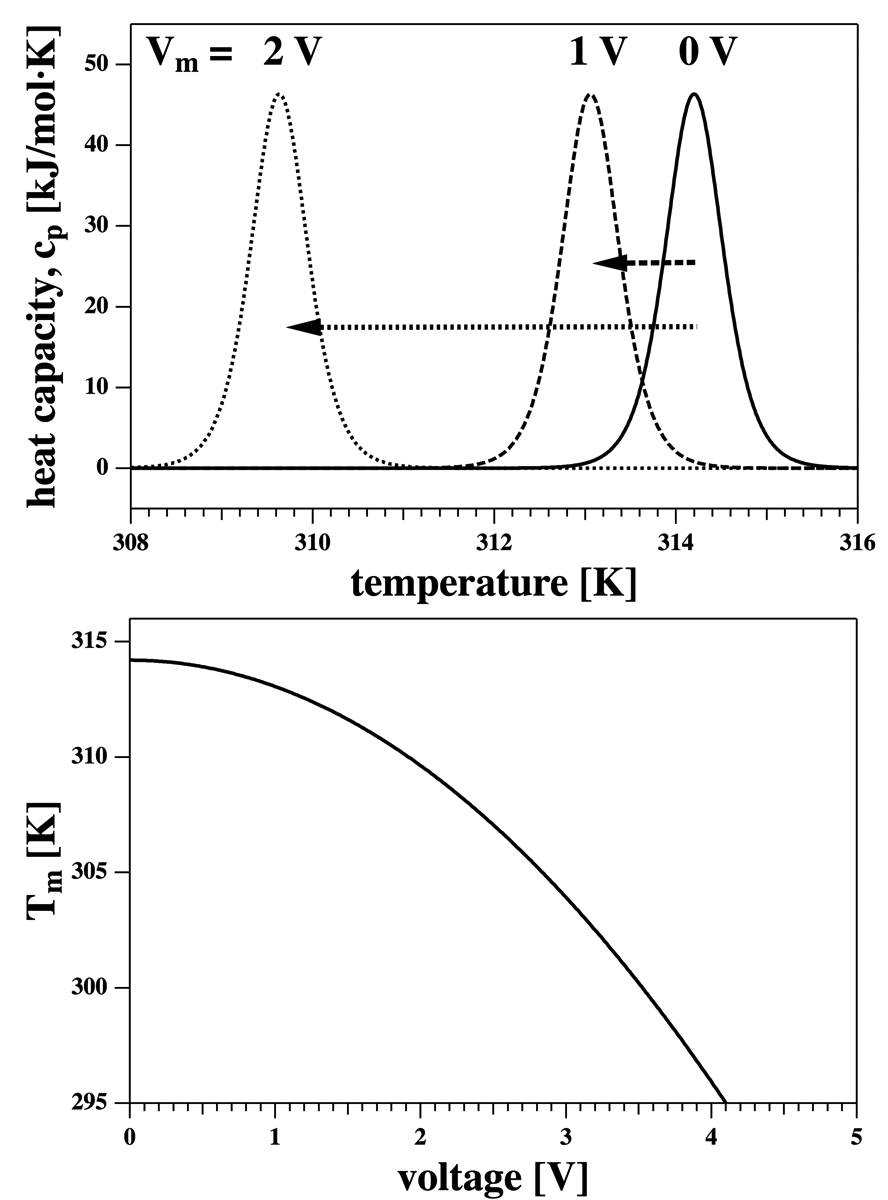}
	\parbox[c]{8.5cm}{\caption{\small\textit{The influence of electrostriction on the melting of membranes. Top: Excess heat capacity of DPPC LUV at three different voltages. Bottom: Change of the transition temperature of DPPC as a function of voltage (see eq. (\ref{eq:T_2.06})).
	}
	\label{Figure1}}}
    \end{center}
\end{figure}
We denote the membrane volume as $V(T)=V_g+\Delta V(T)$ and the membrane area as $A(T)=A_g+\Delta A(T)$, where $V_g$ and $A_g$ are the volume and the area of the gel state lipids, and $\Delta V(T)$ and $\Delta A(T)$ are the temperature dependent changes due to melting. In previous publications we have shown that the changes in both volume and area are proportional to the excess enthalpy during the chain melting transition \cite{Heimburg1998, Ebel2001}:
\begin{eqnarray}
	\label{eq:T_2.02}
	\Delta V(T)=\gamma^V \Delta H_0(T) \;\;\nonumber\\
	\Delta A(T)=\gamma^A \Delta H_0(T) \;,
\end{eqnarray}
where $\gamma^V=7.8\cdot 10^{-10}$m$^3$/J is a constant that is practically independent of the lipid species or the lipid mixture \cite{Ebel2001}, and $\gamma^A=0.89$m$^2$/J. 
We assume that a similar relation holds for the mean thickness,
\begin{equation}
	\label{eq:T_2.03}
	\Delta D(T)=\gamma^D \Delta H_0(T) \;,
\end{equation}
with $\gamma^D=-2.49\cdot 10^{-14}$m/J. The thickness of the gel phase membrane, $D_g$,  is 4.79 nm (for DPPC), and it is $D_f = $ 3.92 nm in the fluid phase. DPPC in the gel phase has an area of $A_g$ = 0.474 nm$^2$ per lipid and a membrane area $1.43\cdot 10^5$ m$^2$ per mol of lipid \cite{Heimburg1998}. Here and below we will assume that the above proportionality to the enthalpy is valid. Further, we assume that the voltage dependence of the pure lipid phases is small.\\
According to eq. (\ref{eq:T_1.02}), the enthalpy change, $\Delta H (T)$, at constant voltage, $V_m$, of the membrane at temperature, $T$, is given by
\begin{eqnarray}
	\label{eq:T_2.04}
	&\Delta H(V_m,T)\!=\!\Delta H_0(T)+\frac{1}{2}\epsilon_0\;V_m^2\int_{D_{g}}^{D(T)}\epsilon\frac{A(T)}{D(T)^2}dD\! =\rightarrow\quad\;& \nonumber\\
	&\Delta H_0(T)\!+\!\frac{1}{2}\epsilon_0\!\;V_m^2\int_{0}^{\Delta H_0(T)}\!\epsilon\frac{\left(A_{g}+\gamma_A\Delta H_0(T)\right)}{\left(D_{g}+\gamma_D \Delta H_0(T)\right)^2}\gamma_D d\Delta H_0(T)\,.&\nonumber\\
\end{eqnarray}
Making use of  $(1+x)^{-2}\approx 1-2x$ for small $x$ and assuming constant $\epsilon$, we finally obtain
\begin{eqnarray}
	\label{eq:T_2.05}
	&&\Delta H(V_m,T)\!=\!\Delta H_0(T)\cdot \rightarrow \\
	&& \left(1\!+\!\frac{1}{2}\epsilon_0\epsilon\;\gamma_D V_m^2\frac{A_{g}}{D_{g}^2} \left[1\!+\!\frac{1}{2}\left(\frac{\gamma_A}{A_{g}}\!-\!2\frac{\gamma_D}{D_{g}}\right)\Delta H_0(T)\right] \right)\;,\nonumber
\end{eqnarray}
where $\Delta H_0(T)$ is the temperature-dependent enthalpy in the absence of voltage described by eq. (\ref{eq:T_2.01}).
One can now determine the temperature dependence of $\Delta H(V_m,T)$. For $T\gg T_m$, $\Delta H(V_m,T)$ assumes a  constant value, which is the excess heat of melting. It is a quadratic function of voltage: $\Delta H(V_m)=\Delta H_0+\alpha_0 V_m^2$ with $\alpha_0=-141.7$ [J/V$^2$] using $\epsilon=4$.\\

$\Delta H_0$ is the melting enthalpy of the membrane in the absence of voltage with melting temperature $T_{m,0}=\Delta H_0/\Delta S_0$, and $\Delta H(V_m)$ is the melting enthalpy of the membrane with voltage $V_m$. The transition temperature $T_m$ in the presence of voltage can be written as
\begin{eqnarray}
	\label{eq:T_2.06}
	T_m&=&\frac{\Delta H(V_m)}{\Delta S_0}=\frac{\Delta H_0 + \alpha_0 V_m^2}{\Delta S_0} =\nonumber\\
	&=&T_{m,0}+\frac{\alpha_0}{\Delta S_0}V_m^2\equiv \left(1+\alpha V_m^2\right) T_{m,0}
\end{eqnarray}
with $\alpha=\alpha_0/\Delta H_0=-0.003634$ [1/V$^2$], and $\Delta S_0=$ \linebreak $ \Delta H_0/T_{m,0}$. The melting profiles and the melting temperature as a function of voltage are shown in Fig.\ref{Figure1}.  One obtains a shift of the transition temperature towards lower temperatures of $-11.4$\,mK for $V_m=100$ mV, of $-1.14$\,K for $V_m=1$\,V and of $-114$\,K for $V_m=$10 V. The effect of electrostriction on the melting temperature is obviously small at physiological voltages. However, it is large for voltages of more than 1 V.\\ 
The above calculation applies to a symmetric membrane. According to the results of \cite{Alvarez1978} the quadratic voltage dependence of $T_m$ can be shifted on the voltage axes when the membrane is asymmetrically charged.  In this case, the maximum melting temperature 
can be achieved for voltages different than zero.

\subsection*{Voltage and temperature dependence of the capacitive susceptibility}\label{Theory_3}
 It has been shown that outside of transitions, capacitance changes induced by voltage are small \cite{White1974, Requena1975b, Farrell2006} but quadratic in voltage. As above we will therefore assume that the voltage dependence of the capacitance in the transition regime is much higher than that of the pure phase. I.e., we assume that $A_g$ and $A_f$ as well as $D_g$ and $D_f$ display a negligible voltage dependence.
\begin{figure}[ht!]
    \begin{center}
	\includegraphics[width=8.5cm]{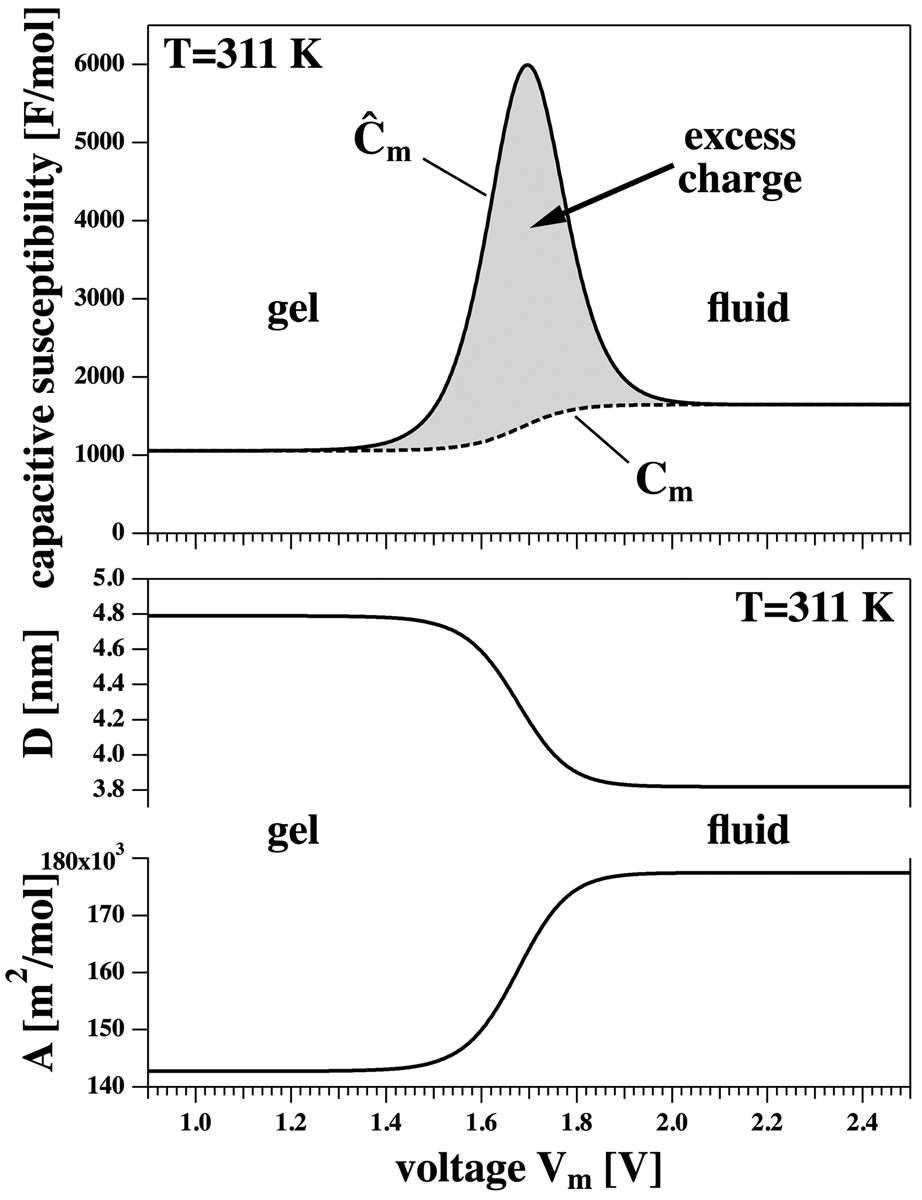}
	\parbox[c]{8.5cm}{\caption{\small\textit{Bottom: Area and thickness changes as a function of voltage at $T=311$\,K.  The temperature, $T=311$ K, is below the melting point of DPPC in the absence of voltage. Top: Capacitive susceptibility, $\hat{C}_m$, of DPPC LUV as a function of voltage. The shaded area indicates the excess charge of the voltage-induced transition. The dashed line is the voltage dependent capacitance, $C_m$. 
	}
	\label{Figure2}}}
    \end{center}
\end{figure}
The charge on a  membrane at temperature $T$ and voltage $V_m$ is then given by
\begin{equation}
	\label{eq:T_3.01}
	q (V_m, T)=\epsilon\epsilon_0 \frac{A_{gel}+\Delta A(V_m,T)}{D_{gel}+\Delta D(V_m,T)}V_m
\end{equation}
and at voltage $V_m+dV$ by
\begin{equation}
	\label{eq:T_3.02}
	q(V_m+dV_m, T)=\epsilon\epsilon_0 \frac{A_{gel}+\Delta A(V_m+dV_m,T)}{D_{gel}+\Delta D(V_m+dV_m,T)}\cdot (V_m+dV_m) \;.
\end{equation}
where $\Delta A$ and $\Delta D$ are again proportional to $\Delta H$ (cf., eq. (\ref{eq:T_2.05})). 
For fixed voltage, the capacitance is a function of temperature only.  For fixed temperature, it is a function of voltage only. 
The area and thickness of the membrane at $T=311$\,K are shown as a function of voltage in Fig. \ref{Figure2} (left).
\begin{figure*}[ht!]
    \begin{center}
	\includegraphics[width=16.5cm]{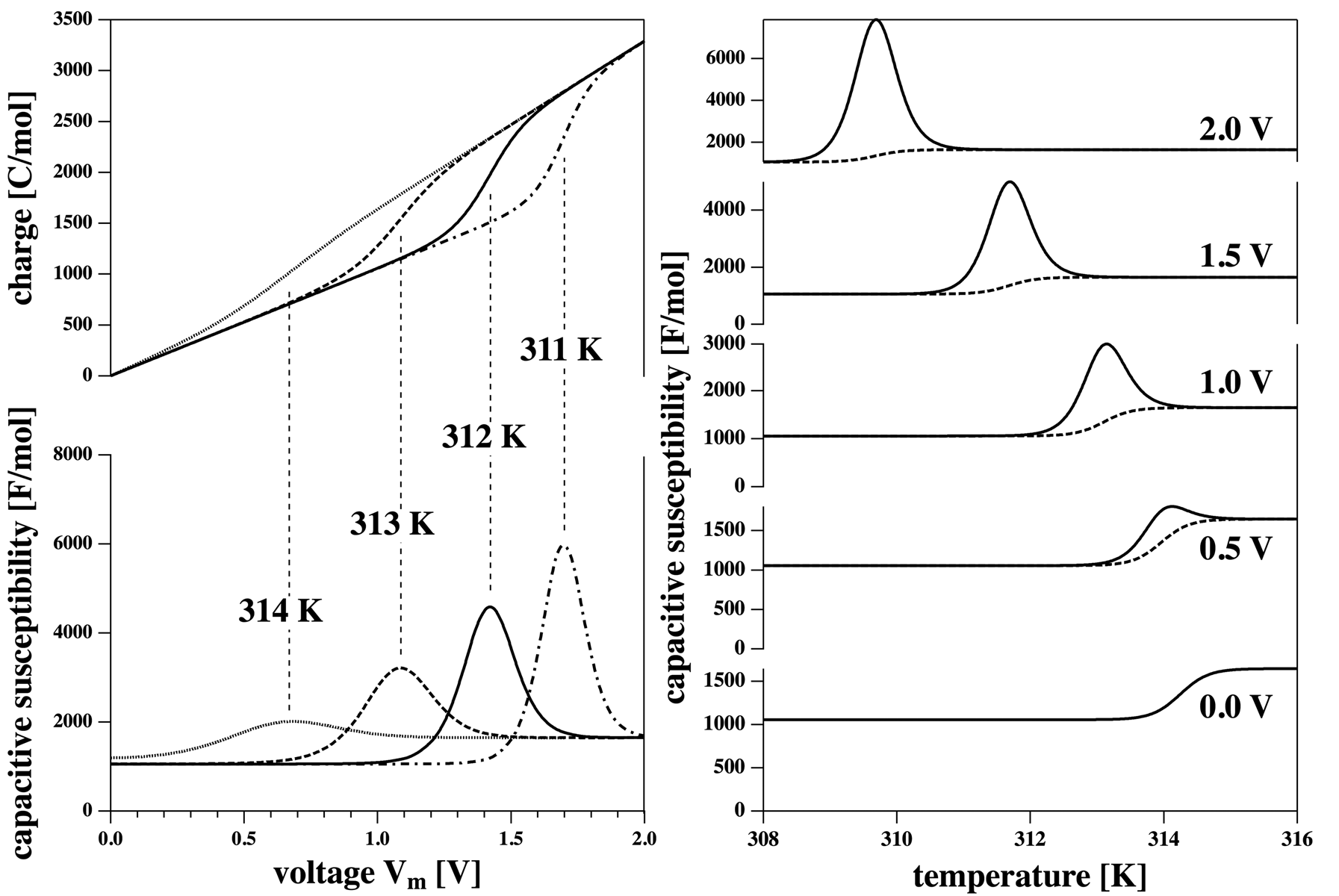}
	\parbox[c]{16cm}{\caption{\small\textit{Left: Voltage-induced transition of DPPC LUV at four different (constant) temperatures. Top panel: the charge as a functions of voltage. Bottom panel: capacitive susceptibility as a function of voltage. Right: Capacitive susceptibility $\hat{C}_m$ as a function of temperature for five different voltages. The dashed lines represent the capacitance $C_m$ as a function of temperature.}
	\label{Figure3}}}
    \end{center}
\end{figure*}
The capacitive susceptibility, $\hat{C}_m$, is now given by
\begin{equation}
	\label{eq:T_3.04}
	\hat{C}_m=\frac{dq}{dV_m}=\frac{q(V_m+dV_m, T)-q(V_m, T)}{dV_m} \;.
\end{equation}
If the temperature $T$ is below the melting temperature, $T_{m,0}$, voltage can induce a transition in the membrane. $\hat{C}_m$ displays a maximum at a transition voltage that depends on the experimental temperature.  This is shown in Fig. \ref{Figure2} (right) for the case of $T=311$K. The units are given in absolute units (per mol of lipid) since the area of the membrane is not constant. For comparison, the specific capacitance in the gel state is $0.74$ $\mu$F/cm$^2$, and in the fluid state it is $0.93$ $\mu$F/cm$^2$.

The charge on the capacitor is given by
\begin{equation}
	\label{eq:T_3.05}
	q(V_m, T)=\int_{V_m}\hat{C}_mdV_m \;.
\end{equation}
The charge as a function of voltage is shown in Fig. \ref{Figure3} (left) for various temperatures. One can see that the charge undergoes a stepwise change at the transition voltage. We call this change in charge the ``excess charge'', $\Delta q_0$. It is given by (cf., eq. (\ref{eq:T.03}))
\begin{equation}
	\label{eq:T_3.06}
	\Delta q_0(T)=\int_{V_m}V_m \left(\frac{\partial C_m}{\partial V_m}\right)dV_m \;,
\end{equation}
where the integral is from a voltage below to a voltage above the transition. The excess charge corresponds to the shaded peak area in Fig. \ref{Figure2} (right). It has a value of 634 C/mol.  

One can also determine the capacitive susceptibility, $\hat{C}_m$, and the capacitance, $C_m$, as functions of temperature. This is shown in Fig. \ref{Figure3} (right) for several voltages. For $V_m=0$, the capacitive susceptibility and the capacitance are identical. 

\subsection*{Fluctuations}\label{Theory_4}
The heat capacity of a membrane at constant pressure is the temperature derivative of the mean enthalpy and is given by 
\begin{equation}
\label{eq:T_4.01}
c_p=\frac{d\left<H\right>}{dT}\quad \mbox{with} \quad \left<H\right>=\frac{\sum_i H_i\exp{(-H_i/kT)}}{\sum_i \exp{(-H_i/kT)}} \;,
\end{equation}
where $\left<H\right>$ is the statistical mean of the enthalpy averaged over all possible microstates of the system with enthalpy $H_i$. 
The enthalpies of the microstates are given by:
\begin{equation}
\label{eq:T_4.02}
H_i=E_i +pV_i+\Pi A_i + \Psi q_i +  ...  
\end{equation}
with the intensive variables pressure $p$, lateral pressure $\Pi$, electrostatic potential $\Psi$ and the conjugated extensive quantities, internal energy $E_i$, volume $V_i$, area $A_i$, and charge $q_i$.

Eq. (\ref{eq:T_4.01}) immediately leads to
\begin{equation}
\label{eq:T_4.03}
c_p=\frac{d\left<H\right>}{dT}=\frac{\left<H^2\right>-\left<H\right>^2}{kT^2} \;;\;  p, \Pi, \Psi, ... =\mbox{const.}
\end{equation}
which is one of the fluctuation relations. 
Similar relations are found for the specific isothermal volume compressibility $\kappa_T^V$, and the area compressibility $\kappa_T^A$ \cite{Heimburg1998}:
\begin{eqnarray}
\label{eq:T_4.04}
\kappa_T^V  =  -\frac{1}{\left<V\right>}\left(\frac{d\left<V\right>}{dp}\right)= \frac{\left<V^2\right>-\left<V\right>^2}{\left<V\right> kT} \;;\; T, \Pi, \Psi =\mbox{const.}&&\\
\kappa_T^A  = -\frac{1}{\left<A\right>}\left(\frac{d\left<A\right>}{d\Pi}\right)= \frac{\left<A^2\right>-\left<A\right>^2}{\left<A\right> kT}\;;\; T, p, \Psi=\mbox{const.}&&
\end{eqnarray}
Like the heat capacity, these compressibilities have maxima in the melting transition \cite{Heimburg1998}. 

The capacitive susceptibility of the membrane, $\hat{C}_m$, is a similar susceptibility and can be written as 
\begin{equation}
\label{eq:T_4.05}
\hat{C}_m=- \frac{dq}{d\Psi}=\frac{dq}{dV_m}=\frac{\left<q^2\right>-\left<q\right>^2}{kT}\;;\; T, p, \Pi =\mbox{const.}
\end{equation}

All of the above susceptibilities are derivatives of extensive variables with respect to the conjugated intensive variables. For such susceptibilities, the fluctuations $\left<X^2\right>-\left<X\right>^2$ are quadratic forms and therefore always positive. Heat capacity, volume and area compressibility, and capacitance must always be positive definite functions. For this reason, one also finds that the integrals of the susceptibilities are always be positive:
\begin{eqnarray}
\label{eq:T_4.06}
\Delta H &=\int_{T_1}^{T^2} c_p dT > 0 \ \ \ \ \ \qquad & \mbox{for}\qquad T_2>T_1\nonumber\\
\Delta V &= -\int_{p_1}^{p_2 }\left<V\right>\kappa_T^V dp > 0 \qquad & \mbox{for}\qquad p_2>p_1\nonumber\\
\Delta A &= -\int_{\Pi_1}^{\Pi_2 }\left<A\right>\kappa_T^A d\Pi > 0 \qquad & \mbox{for}\qquad \Pi_2>\Pi_1\\
\Delta q &=\int_{V_{m,1}}^{V_{m,2}} \hat{C}_m dV_m > 0 \qquad & \mbox{for}\qquad V_{m,2}>V_{m,1} \;.\nonumber
\end{eqnarray}
This implies that an increase in voltage across a membrane must result in an increase in charge so long as intensive variables  other than $V_m$ are kept constant.  This is in agreement with the findings in Fig. \ref{Figure3} (left).

For derivatives of extensive quantities with respect to non-conjugated quantities, the fluctuations are no longer positive definite forms, and negative values can be obtained.  For instance, the volume expansion coefficient is given by $d\left<V\right>/dT$\- =$\left[\left<VH\right>-\left<V\right>\left<H\right>\right]/kT^2$. For water at $0^{\circ}$C it is negative. Similarly, derivatives of other extensive variables with respect to non-conjugated intensive variables, such  as $d\left<q\right>/dT$ or \linebreak $d\left<q\right>/d\Pi$,  can also be negative (see below). 

\subsection*{Piezoelectricity}\label{Theory_5}
A material is said to be ``piezoelectric" if the application of a force produces an electric field (and vice vera). ``Piezo'' originates from the Greek word for pressure, and we will therefore use the term `piezoelectric' as synonymous with `electromechanical' in the sense of pressure-induced voltages across membranes. At fixed temperature,
\begin{equation}
	\label{eq:T_5.01}
	dq=\left(\frac{\partial q}{\partial V_m}\right)_{\mathcal{F}}dV_m + \left(\frac{\partial q}{\partial \mathcal{F}}\right)_{V_m}d\mathcal{F} \;,
\end{equation}
where $\mathcal{F}$ is the force normal to the membrane. Since thickness changes in the melting transition are coupled to area changes, this leads to the relation
\begin{eqnarray}
	\label{eq:T_5.02}
	dq&=&\left(\frac{\partial q}{\partial V_m}\right)_{\mathcal{F}}dV_m + \left(\frac{\partial q}{\partial \Pi}\right)_{V_m}\underbrace{\left(\frac{\partial \Pi}{\partial \mathcal{F}}\right)_{V_m}d\mathcal{F}}_{d\Pi} \nonumber\\
	&=& \left(\frac{\partial q}{\partial V_m}\right)_{\Pi}dV_m + \left(\frac{\partial q}{\partial \Pi}\right)_{V_m}d\Pi  \;,
\end{eqnarray}
where $\Pi$ is the lateral pressure of the membrane. In the following we will focus on lateral pressure changes.
\begin{figure}[ht!]
    \begin{center}
	\includegraphics[width=8.5cm]{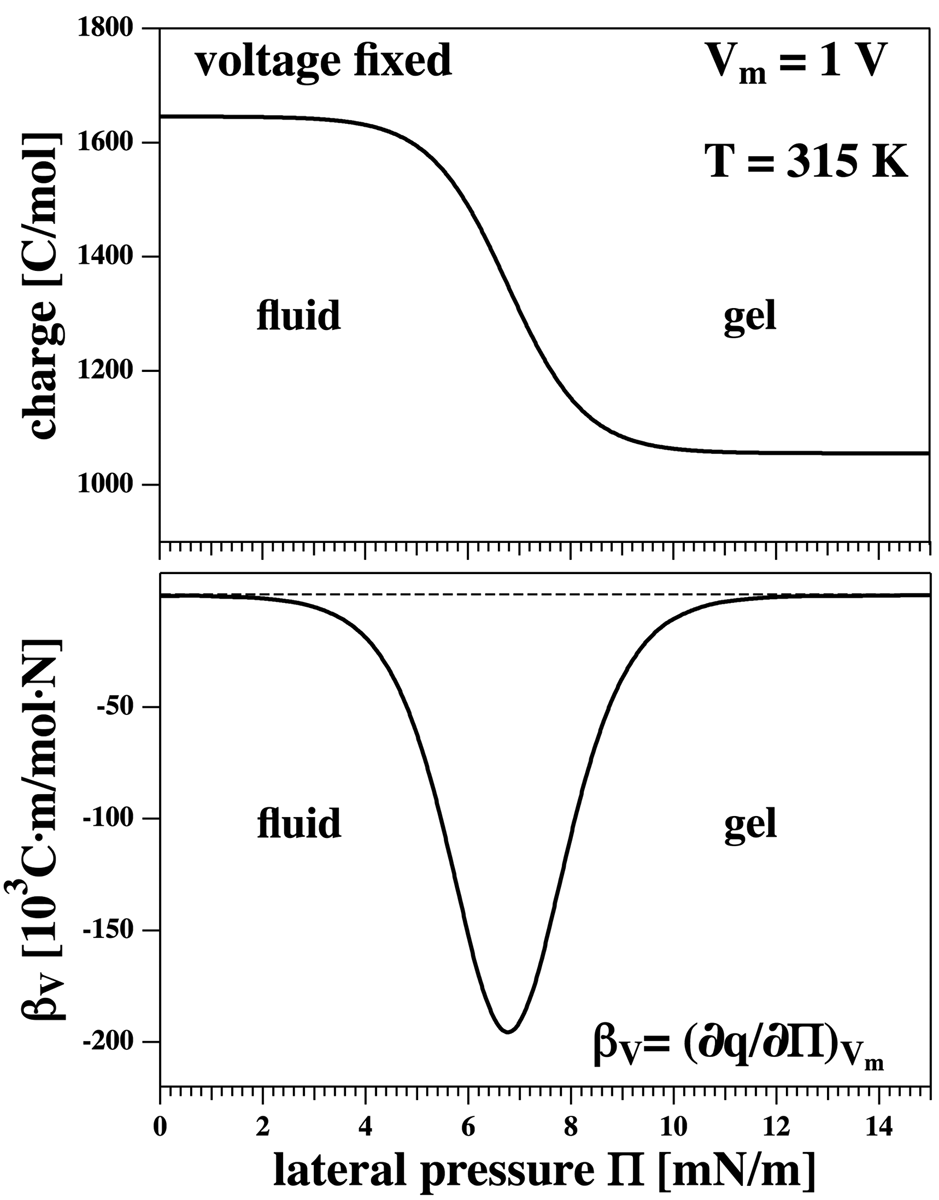}
	\parbox[c]{8.5cm}{\caption{\small\textit{Changes in electrical properties induced by lateral pressure changes. Top: Change in the charge on a capacitor upon changes in lateral pressure at fixed voltage of $V_m=$1\,V. Bottom: Change in voltage induced by lateral pressure changes at fixed charge on the capacitor ($C_m$=1430.7 C/mol corresponding to the charge on the fluid phase membrane at $V_m$=1\,V). Increasing lateral pressure renders the membrane more solid.
	}
	\label{Figure4}}}
    \end{center}
\end{figure}
At constant voltage, the enthalpy change, $\Delta H (T)$, of the membrane at temperature $T$ due to a lateral pressure is given by a modified version of eq. (\ref{eq:T_2.04}):
\begin{eqnarray}
	\label{eq:T_5.04}
	\Delta H(V_m,T,\Pi)&=&\Delta H_0(T)+\Delta W_c(V_m)+\Delta W_A(\Pi)\nonumber\\
	\Delta W_A(\Pi)&=& \Pi\Delta A=\Pi \gamma^A \Delta H_0(T)\;,
\end{eqnarray}
where $\Delta W_A(\Pi)$ is the work done to change the area of the membrane from $A_g$ to A. 
With the help of eq. (\ref{eq:T_2.05}) this leads to
\begin{eqnarray}
	\label{eq:T_5.05}
	&&\Delta H(V_m,T,\Pi)= \Delta H_0(T)(1+\frac{1}{2}\epsilon_0\epsilon\;\gamma_D V_m^2\frac{A_{gel}}{D_{gel}^2}\cdot \rightarrow \nonumber\\
	&&\left[1-\frac{1}{2}\left(\frac{\gamma_A}{A_{gel}}
	+2\frac{\gamma_D}{D_{gel}}\Delta H_0(T) \right)\right]+\gamma^A\Pi ) \, ,
\end{eqnarray}
and the melting temperature is given by
\begin{equation}
	\label{eq:T_5.06}
	T_m=\frac{\Delta H(V_m, \Pi)}{\Delta S_0}=\left(1+\alpha V_m^2  + \gamma^A \Pi \right)\,T_{m,0}
\end{equation}
with $\alpha=-0.003634$ [1/V$^2$] (see eq. (\ref{eq:T_2.06}), and $\gamma^A=0.89$ m$^2$/J. 

The charge on a  membrane at temperature T, voltage $V_m$ and lateral pressure $\Pi$ can be written as
\begin{equation}
\label{eq:T_5.07}
q (V_m, T, \Pi)=\epsilon\epsilon_0 \frac{A_{gel}+\Delta A(V_m,T, \Pi)}{D_{gel}+\Delta D(V_m,T,\Pi)}V_m
\end{equation}
where $\Delta A$ and $\Delta D$ are again proportional to $\Delta H$ (cf. eq. (\ref{eq:T_5.05})) 

Now we can determine (1) how the charge on the capacitor changes with changes in lateral pressure at constant voltage and temperature 
(and vice versa) and (2) how the voltage changes at constant temperature and constant charge with changes in lateral pressure (and vice versa). 
\begin{figure}[ht!]
    \begin{center}
	\includegraphics[width=8.5cm]{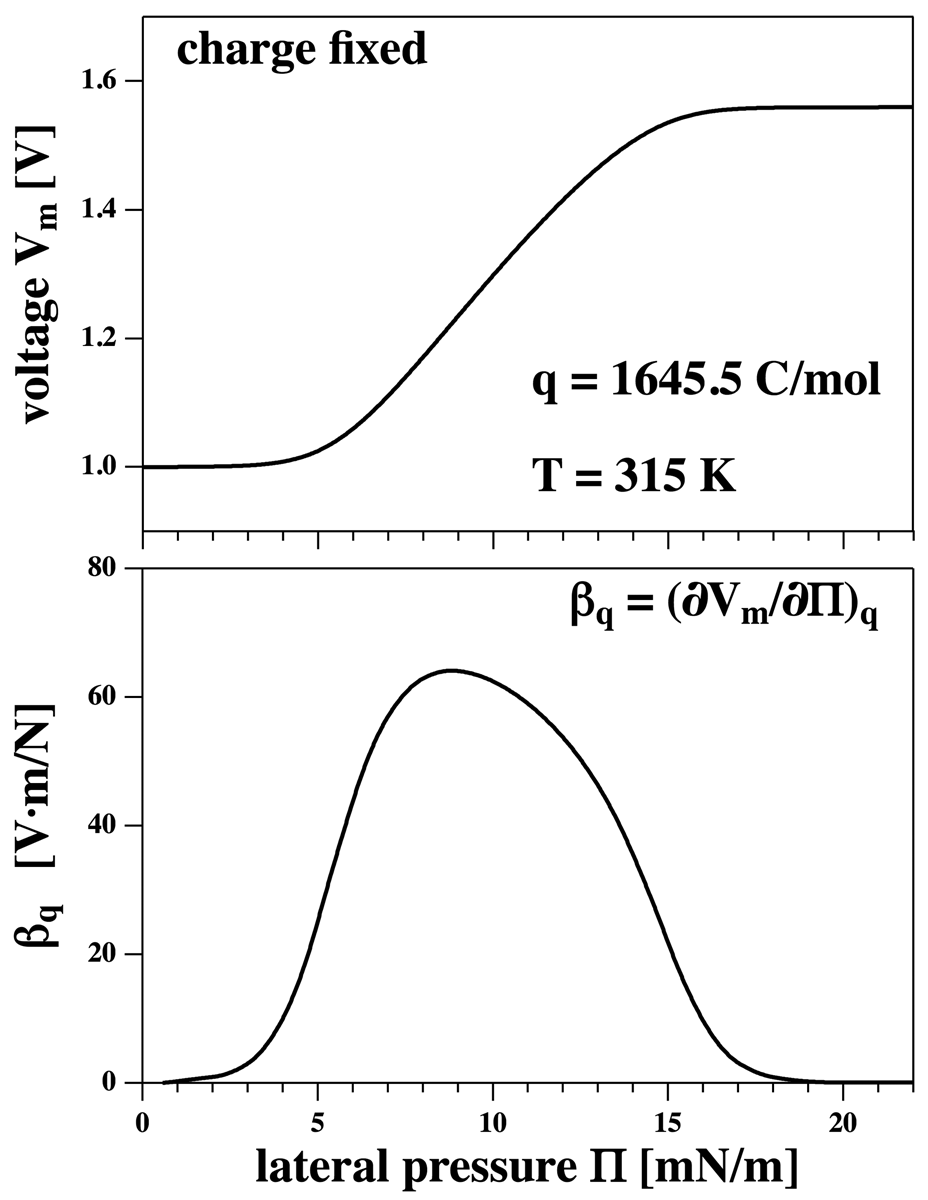}
	\parbox[c]{8.5cm}{\caption{\small\textit{Top: Change in voltage induced by lateral pressure changes at fixed charge on the capacitor ($C_m$=1430.7 C/mol corresponding to the charge on the fluid phase membrane at $V_m$=1\,V). Increasing lateral pressure renders the membrane more solid. Bottom: The corresponding susceptibility, $\beta_q=\left(\partial V_m/\partial \Pi\right)_q$.
	}
	\label{Figure5}}}
    \end{center}
\end{figure}

Fig. \ref{Figure4} (left) shows case 1 for a fixed voltage of $V_m$=\,1\,V and $T$=\,315\,K. Under these conditions, the membrane is in the fluid state when the lateral pressure $\Pi$ is zero. Increasing pressure renders the membrane more solid and the capacitance of the membrane decreases. This leads to a release of charge from the capacitor (i.e., pressure induced capacitive currents). One can define the corresponding susceptibility
\begin{equation}
	\label{eq:T_5.09}
	\beta_V\equiv \left(\frac{\partial q}{\partial \Pi}\right)_{V_m} \;,
\end{equation}
which is shown in Fig. \ref{Figure4} (right). This susceptibility represents the change in the charge on the membrane due to an increment in the pressure at constant voltage. Since it is a derivative of the extensive variable $q$ with respect to the non-conjugated extensive variable $\Pi$, it 
can have a negative value.

When the charge in eq. (\ref{eq:T_5.01}) is kept constant ($dq=0$), we obtain
\begin{equation}
\label{eq:T_5.03}
	dV_m=\left[-\frac{\left(\frac{\partial q}{\partial \Pi}\right)_{V_m}}{\left(\frac{\partial q}{\partial V_m}\right)_{\Pi}}\right]d\Pi \equiv \beta_q d\Pi \;.
\end{equation}
This is shown in Fig. \ref{Figure5} (left) where the charge is fixed at $q=$1645.4 C/mol, which is the charge on the fluid membrane at a voltage of $V_m=$\,1\,V. The work done on the membrane by pressure is converted into work done on the charges by changing their distance. Increasing pressure at fixed charge leads to a larger voltage across the membrane. We call this a piezoelectric effect. The coupling constant, 
\begin{equation}
	\label{eq:T_5.10}
	\beta_q=\left(\frac{\partial V_m}{\partial \Pi}\right)_q
\end{equation}
is shown in Fig. \ref{Figure5} (right). It is identical to the term in rectangular brackets in eq. (\ref{eq:T_5.03}). A situation of constant charge could be present in a membrane during a sudden reversible change in lateral pressure when charges have no time to dissociate.  


\section*{Discussion}
Lipid membranes are thin capacitors that are unique in their property to display transitions in capacitance. Here, we have offered 
a theoretical framework for describing the capacitance and the capacitive susceptibility of membranes in the transition regime. Since biomembranes are close to such transitions, this is of immediate biological relevance. Charges on membranes create forces on the membrane that tend to compress the membrane normal to the membrane surface. This effect is called electrostriction  \cite{Vanselow1966a}. It leads to a lowering of the melting temperature of the membrane. Therefore, a change of transmembrane voltage can induces melting transitions. As a consequence, there is an excess charge linked to the voltage-induced transition. This is expressed by the capacitive susceptibility, $\hat{C}_m=\partial q/\partial V_m$, which displays a pronounced maximum at the transition.  This behaviour is very similar to that of the heat capacity as a function of temperature and the volume and area compressibilities as a function of hydrostatic and lateral pressure. These functions are all derivatives of  extensive thermodynamic variables with respect to their conjugated intensive variable. Using the fluctuation theorem, we have shown that the capacitive susceptibility is related to fluctuations in charge and must therefore always be positive. The maximum of the capacitive susceptibility describes the fact that in particular voltage regimes close to transitions a small change in voltage may lead to a large uptake of charge, i.e., the membrane is very susceptible to small variations in conditions. 

We further show that there exist electromechanical couplings in lipid membranes. One finds a straight-forward connection between lateral pressure and the charge on a membrane (at constant voltage) that also assumes a maximum in the melting transition of the membrane. This implies that changes in pressure can give rise to capacitive currents. This effect is described by the derivative of an extensive variable (charge) with respect to a non-conjugated variable (lateral pressure). Such susceptibilities may have either positive and negative values. At constant charge, there also exists a coupling between voltage and lateral pressure. This  effect corresponds to piezoelectricity. The coupling constant assumes a maximum in the melting transition. This effect is very important for excitatory processes (for instance nerve pulse) and is discussed below. In the present paper, we have not considered other potentially significant effects such as field-induced changes in polarization. Such effects are relevant and clearly exist in lipid monolayers. This will be the focus of a future publication. 

Due to their different capacitance, gel and fluid membranes carry different charges when a constant voltage is applied. An interesting consequence is that the presence of a transmembrane voltage will lead to charge separation in the melting regi\-me where gel and fluid domains coexists. If the membranes contain a fraction of charged lipids, this will lead to an asymmetric accumulation of charged lipids in the fluid domains. Thus, the shape of phase diagrams will be influenced by the application of an electrical field across the membrane. Not surprisingly, besides temperature, pressure and the concentrations of the components, the transmembrane voltage is also a variable that determines phase diagrams and should be included in Gibbs' phase rule.

The influence of voltage on the capacitance of black lipid membranes had been investigated both theoretically and experimentally in several studies in the 1970's \cite{White1970, White1973, White1974, White1981, Alvarez1978}. All of these studies assume that electrostriction is the dominant effect leading to a reduction in membrane thickness and an increase in area due to an increase in voltage. Those studies were done far away from the phase transitions temperature of black lipid membranes.  In agreement with our derivations here, they found that the dependence of the capacitance should be a quadratic function of voltage, i.e., $C_m\propto (1+ \alpha V_m^2)$, where $\alpha$ is a constant. This constant is strongly influenced by the presence of solvent in the black lipid membranes \cite{Requena1975b}. Solvent-free membranes display a much smaller voltage dependence of the capacitance on voltage. In an interesting experimental study, Alvarez and Latorre \cite{Alvarez1978} found that changes in capacitance in asymmetric membranes are influenced by a resting potential, $E_0$, such that $C_m\propto (1+\alpha (V-E_0)^2)$. Alvarez and Latorre discussed this finding in the context of nerve pulse propagation and the measurement of gating currents.  They suggested that the membrane itself could display capacitive currents similar to gating currents. The novel aspect of the present paper is the recognition of the profound effect that the melting transition has on the nonlinear behavior of the membrane capacitance. While previous authors have assumed a constant compressibility of the membrane, we have made use of the fact that the compressibility is dramatically increased in transitions. 

The two terms of the capacitive current $C_m\cdot dV_m/dt$ and $V_m\cdot d C_m/dt$ in eq. (\ref{eq:Intro.03}) can have different time dependences. The first term is fast and is largely determined by the electrical resistance of the aqueous medium. The second term has a time scale given by the relaxation time of the membrane, which is slow in transitions. It should therefore be possible to distinguish the two contributions to the capacitive current experimentally. We have previously shown that relaxation time scales can be as large as 30 seconds for artificial membranes at the transition maximum, and we estimate time scales of up to 100ms for biomembranes \cite{Grabitz2002, Seeger2007}. Besch and collaborators investigated HEK293 cells and found currents with timescales in the 10ms regime after voltage change \cite{Besch2003}, which indicates that these currents are related to time-dependent changes in membrane geometry rather than being caused by charging a membrane with a constant capacitance. Slow capacitive currents after voltage changes were also found in rat nerves \cite{Kilic2001} but interpreted as gating currents.

The electromechanical aspect of membranes has received considerable interest in the past. Ochs and Burton showed in 1974 that black lipid membranes made of egg-PC and cholesterol show electromechanical behavior under voltage clamp conditions \cite{Ochs1974}. They applied an oscillating pressure difference across the membrane and recorded capacitive currents that they described by $I_C=V_m\cdot dC_m/dt$ (cf., eq. (\ref{eq:Intro.03})). 

As mentioned above, the interesting issue of polarization has not been treated here.  However, much of the literature on electromechanical phenomena in membranes is dedicated to polarization caused by membrane curvature. R. B. Meyer proposed in 1969 that liquid crystals of molecules that possess electrical dipole moments should also be piezoelectric \cite{Meyer1969}. In particular he stated that ``another possible application of these effects may be in interpreting some of the potentials and ion distributions observed in liquid-crystalline biological structures.'' Following this idea, Petrov and collaborators proposed that membranes should also display piezoelectric behavior \cite{Petrov1984}. Lipid monolayers have significant dipole moments due to the polar head groups and the oriented associated water layer. For a symmetric membrane of a zwitterionic lipid, the polarizations of the two layers should cancel because they have opposite orientations. However, in the presence of curvature, this symmetry argument does not apply, and one expects an electrical field generated by curved membranes. Petrov named this phenomenon ``flexoelectricity'' and demonstrated the effect in experiments similar to those by Ochs and Burton \cite{Ochs1974} but interpreted differently as curvature-induced polarization \cite{Petrov1989, Petrov1994}. In several further papers the authors applied the concept of flexoelectricity to biomembranes and proposed a coupling of flexoelectricity to ionic currents through channel proteins \cite{Petrov1993, Petrov1997}. Since lipid membranes in the absence of proteins also display ion-channel-like characteristics close to transitions \cite{Blicher2009, Heimburg2010, Laub2012}, Petrov's considerations are also valid here.  It is reasonable to expect that the polarization, $P$, of membranes changes significantly in the phase transition regime and that the related 
susceptibility, $dP/dE$ (with E being the electrical field), should have a maximum in the transition. This must be so since the dipole moment of a gel monolayer differs from that of a fluid monolayer. Upon membrane bending one should find an asymmetric distribution of gel and fluid lipids in the two opposing monolayers \cite{Ivanova2001} leading to an effective polarization of the membrane. Unfortunately, we are not aware of any reliable experimental data or theoretical estimates of the magnitude of this polarization.  To our knowledge, the above authors did not investigate the effect of the phase transition on flexoelectricity. However, polarizations induced by curvature are completely analogous to voltage changes induced by lateral compression as described above by eq. (\ref{eq:T_5.03}) and Fig. \ref{Figure5}. Interestingly, Helfrich investigated the effect of voltage on the phase transition temperature of 3-dimensional liquid crystals as early as 1970  \cite{Helfrich1970}.  He found that the shift of the transition is given by $\Delta T_m=\frac{1}{2} T_{m,0}\epsilon_0\Delta \epsilon E^2/\Delta H_0 \rho$, where $\Delta \epsilon$ is the difference of the dielectric constant between liquid and solid phase, $\rho$ is is the density and $E$ is the electric field. Assuming $E/D =V_m$, this law is analogous to eq. \ref{eq:T_2.06}. 

It is known experimentally that lipid monolayers have large dipole potentials of order 300mV --- somewhat higher for gel than for fluid phase monofilms \cite{Vogel1988, Lee1995}. This is frequently attributed to the dipolar nature of the lipid head group and the associated water. One can influence the state of lipid monolayers in experiments by an applied field.  At positive voltages a liquid-expanded monolayer becomes more solid, while the opposite effect is observed when the field is reversed. This rules out the possibility that the influence of voltage is due to electrostriction (PhD Kasper Feld, NBI 2012), which is independent of the direction of the field.  For membranes, this is less clear. Antonov and collaborators \cite{Antonov1990} measured the voltage dependence of the lipid chain melting transition via the effect of voltage on membrane permeability, which 
displays a maximum at the melting point.  For synthetic black lipid membranes made of either DPPC or DPPA (dipalmitoyl phosphatidic acid), they found an increase  in melting temperature that was well described by
\begin{eqnarray}
\label{eq:T2.1_01}
T_m(\mbox{DPPC}) & = & 315.4 [K] + 20.5 [K/V]\cdot V_m \;,\nonumber \\
T_m(\mbox{DPPA}) & = & 332.8 [K]  +55.1 [K/V]\cdot V_m\;.
\end{eqnarray}
This corresponds to a linear shift of $T_m$ of +1.03 K at $V_m=50$mV for DPPC and a shift of 2.75 K for DPPA towards higher temperatures, respectively. This effect is opposed to the trend towards lower temperatures predicted above based on electrostriction, which is in agreement with previous predictions from Sugar \cite{Sugar1979}. However, Antonov's finding is remarkably close to the calculation of Cotterill \cite{Cotterill1978} made by  considering the polarization of the monolayers. The shift of $T_m$ by Antonov was found to be linear within experimental accuracy as expected from Cotterill's calculation.  We do not believe, however, that the derivation of Cotterill is theoretically sound.  \\
An increase of the transition temperature with increasing voltage poses an interesting theoretical problem. Under such circumstances, the fluid membrane can be made solid by voltage leading to a lower capacitance. This would render the excess capacitive susceptibility in the transition negative an would result in a negative excess charge. According to the considerations of fluctuations above, this can hold only if there are couplings between the charge on the membrane and non-conjugated intensive variables. It remains to be seen whether the simultaneous change of capacitance and polarization allows for such behavior. However, if this were the case, an increase in voltage could possibly result in a decrease of charge, i.e., in capacitive currents against the applied field. It should also be noted that the publication of Antonov and collaborators \cite{Antonov1990} is the only one on voltage-induced shifts in transition temperature that we are aware of, and independent experiments verification would be useful.  In this respect, the shift of the quadratic dependence on voltage due to a pre-existing polarization found by Alvarez and Latorre \cite{Alvarez1978} might contain the answer to the present problem. This suggests that the membranes in the experiments of Antonov were not symmetric.  Additional studies of polarized membranes are to be encouraged. 

The biological importance of electromechanical coupling was  discussed in connection with the function of the hair cells of the outer ear \cite{Rabbitt2005, Sachs2009}. In a recent paper,  Brownell and collaborators showed that tethers pulled from hair cells contract upon application of a voltage difference across the cell body \cite{Brownell2010}. The phenomena discussed in the present publication are especially important for nerve pulse propagation. A change of the membrane state from fluid to gel will alter the capacitance by about 50\%,see eq. (\ref{eq:T_1.05}). If this change occurs within 0.5 ms (which is the time scale of the rising phase of the nervous impulse), it will generate a significant capacitive current of order 100 $\mu$A/cm$^2$ at a constant voltage of $V_m =$ 100 mV.  Ionic currents of a similar order of magnitude are central elements in the Hodgkin-Huxley model of the nerve pulse \cite{Hodgkin1952b} (Fig. 18 therein). Hodgkin and Huxley calculated net ion currents of the order of 100-600 $\mu$A/cm$^2$ in the squid axon. In previous publications we have suggested that the nervous impulse consists of an electromechanical soliton corresponding to a lateral compression of the neuronal membranes \cite{Heimburg2005c, Heimburg2007b, Heimburg2008, Andersen2009, Villagran2011, Lautrup2011}. In particular, we have proposed that the membrane undergoes a change in state from fluid to gel and back during the nerve pulse. Thus, in the soliton theory one expects capacitive currents of similar magnitude as the ionic currents in the Hodgkin-Huxley model. If charges cannot dissociate (i.e., no capacitive current), one rather expects changes in transmembrane potential (Fig. \ref{Figure5}).  The consequence of these non-linear effects is an electromechanical soliton that can travel along membrane cylinder with a velocity close to the speed of sound with many similarities to the nervous impulse. In fact, Tasaki and collaborators showed in various publications that such mechanical pulses exist \cite{Iwasa1980a, Iwasa1980b, Tasaki1980, Tasaki1982b, Tasaki1989, Tasaki1990}. Further support comes from atomic force experiments, that show mechanical signals in synapse in phase and proportional to voltage changes \cite{Kim2007}. Further, light scattering techniques demonstrate that voltage changes are accompanied by changes in nerve dimensions \cite{Tasaki1968, Rector2009}. 

One might reasonably assume that polarization pulses can also propagate in membranes. In lipid monolayers, such pressure pulses have recently been demonstrated experimentally by Schneider and collaborators \cite{Griesbauer2012}. The pressure pulses are accompanied by voltage pulses that are directly related to and in phase with the pressure pulse (M. F. Schneider, private communication). The coupling displays a maximum in the phase transition regime in an agreement with the concepts proposed here. Such experiments are important and will eventually lead to a complete thermodynamic picture of the capacitive susceptibility and the electromechanical behavior of biomembranes and nerves.

\section*{Conclusion}
We have shown here that lipid membranes are very susceptible to voltage changes close to phase transitions. This result extends previous theoretical and experimental findings of voltage-dependent capacitances of artificial membranes far from such transitions. We have introduced a capacitive susceptibility that displays a pronounced maximum at the transition. Voltage chan\-ges generate changes in area that result in an electromechanical coupling.  Since biomembranes are exist naturally in a state close to a transition, this effect will play a role in excitatory processes of the cell. One important example is nerve pulse propagation.\\


\noindent\textbf{Acknowledgments:} I acknowledge communication with Dr. K. Vanselow from Kiel, who explored the influence of voltage on the mechanics of membranes as early as 1963. I thank Prof. A. D. Jackson from the NBI for critical reading and helpful comments. The support of the Villum Foundation (Denmark) is gratefully acknowledged.


\footnotesize{

}

\end{document}